# Will recently proposed experiments be able to demonstrate quantum behavior of entire living organisms?

## C. L. Herzenberg

**Abstract**

Recently proposed experiments consider creating and observing the quantum superposition of small living organisms. Those proposed experiments are examined here for feasibility on the basis of results of earlier studies identifying a boundary separating obligatory classical behavior from quantum behavior. It appears that the proposed experiments may be expected to succeed for the case of viruses, but most probably fail for the case of the appreciably larger organisms that are also considered.

## Introduction

The existence of superposed quantum states has been demonstrated for a variety of small physical objects, including electrons, atoms, ions, and molecules, and also including larger molecules such as fullerene buckyballs. A recent article proposes setting up experiments to attempt to create and examine quantum superposition of larger objects, notably including small organisms.[1] For how large an object or organism would this be demonstrable?

A number of studies have examined aspects of the behavior of quantum objects in an expanding universe of limited duration.[2-8] These studies have suggested that quantum objects in an expanding universe must necessarily experience a limitation on the spatial extent of their quantum wave structure, and it follows from these results that sufficiently large objects must exhibit classical rather than quantum behavior.[2-8] The boundary between obligatory classical behavior based on these effects and allowed quantum behavior can be expressed in terms of size and mass or in terms of the moment of inertia. Thus, from an estimate of the magnitude of the moment of inertia of an object, it is possible to arrive at a conclusion as to whether a particular object would be obligatorily classical due to these effects; or whether it might instead exhibit quantum behavior under suitable conditions. We can use this criterion to provide guidance as to whether the organisms proposed for study by Romero-Isard and his colleagues might be able to exhibit quantum superposition as entire objects, or whether instead they would not exhibit superposition, and behave instead as classical objects.

## Analysis and discussion

We address the question of whether Romero-Isart and his colleagues Juan, Quidant, and Cirac may be expected to be able to demonstrate quantum superposition in any of their proposed target objects or organisms, or whether the presence of a fundamental constraint might forbid quantum superposition in any of the proposed experiments.[1]

These authors propose an experiment in which an attempt would be made to place a target organism such as a virus into a quantum superposition of states by conducting quantum optomechanical experiments. The organism would be maintained in a low vacuum in isolation from the environment by levitating it inside an optical cavity using an optical tweezers technique or by self-trapping using two cavity modes.[1] Because the organism is not in direct thermal contact with a substrate, it is considered possible to use laser Doppler cooling to cool the organism down to the ground state of its mechanical motion.[1] The organism might then be excited into a resonant quantum excited state of mechanical motion, and consequently might also potentially be excited into a superposition of the ground and the excited state. Romero-Isart and colleagues analyse the possibility of performing the proposed experiment with living organisms, and indicate that the size of some of the smallest living organisms, such as spores and viruses, is comparable to the laser wavelength considered for their experiment, as required.[1] These organisms would have to survive the vacuum, behave optically like dielectric objects, and be largely transparent to the laser light, issues that have all been addressed by the authors. But, beyond these and other experimental issues, are there fundamental issues that could interfere with such an experiment? Could this quantum superposition of states take place, and could it be observed?

Possible inherent limitations on quantum behavior for objects in an expanding universe of finite duration have been studied earlier.[2-5] This work has shown that sufficiently large objects in an expanding universe might be expected to behave classically, while small objects would exhibit quantum behavior.[2-5] The threshold between classical and quantum behavior, i.e. the quantum-classical boundary, is sensitive to both the size and the mass of an object, and turns out to depend directly on the magnitude of the moment of inertia.[6-8] The threshold moment of inertia has been estimated to be given approximately by the equation:[6,8]

$$I_{th} \approx h/4\pi H_o \qquad (1)$$

Here, h is Planck's constant, and $H_o$ is the Hubble constant. It should be noted that this criterion can at best be expected to provide only a very rough estimate of a threshold separating obligatory classical from possible quantum behavior of physical objects.

Eqn. (1) provides a very straightforward criterion for a boundary separating the smaller objects potentially exhibiting quantum behavior from those larger objects that apparently will necessarily exhibit classical behavior as a result of effects associated with properties of the universe such as its duration and expansion rate.[6,8]

We can evaluate this threshold numerically; we will use $h = 6.63 \times 10^{-34}$ joule-seconds as the value for Planck's constant, and $H_o = 2.3 \times 10^{-18} \sec^{-1}$ as the value for the Hubble constant. Inserting these values into Eqn. (1), we can evaluate a numerical value for the threshold moment of inertia in mks or SI units as:

$$I_{th} \approx 2.3 \times 10^{-17} \text{ kg·m}^2 \qquad (2)$$

This result tells us that, approximately speaking, any object with a moment of inertia larger than about $10^{-17}$ kg·m$^2$ would be expected to behave in a classical manner, while any object with a moment of inertia smaller than about $10^{-17}$ kg·m$^2$ may exhibit quantum behavior, unless brought into classical behavior by other effects such as quantum decoherence.[9-11]

Let's look at the magnitudes of the moment of inertia of some objects that have previously been shown to behave quantum mechanically as entire objects and the magnitudes of the moments of inertia of the small organisms under discussion, and compare them with the threshold moment of inertia given in Eqn. (2).

We know that some small physical objects such as electrons, atoms, and small molecules can behave quantum mechanically as entire objects. Small molecules have relatively small moments of inertia. As an example, the moments of inertia for a water molecule with respect to different axes through the center of mass are reported in the range of about $1 \times 10^{-47}$ kg·m$^2$ to $3 \times 10^{-47}$ kg·m$^2$.[12] These moment of inertia values are some 30 orders of magnitude smaller than the critical threshold moment of inertia evaluated above in Eqn. (2), and thus are well within the range of expected quantum behavior according to this criterion.

The largest molecules for which successful quantum interference experiments have been reported are medium-sized molecules, the fullerenes.[13-15] For orientation purposes, the diameter of a $C_{60}$ fullerene buckyball is about a nanometer. Successful quantum interference experiments with fullerenes (both $C_{60}$ and $C_{70}$ molecules) have been carried out.[9,13] Research groups have sent fullerene molecules with 60 or 70 carbon atoms each through the equivalent of two-slit interference equipment, dramatically displaying their quantum wave nature as entire objects in translational motion. Those quantum interference experiments have established clearly that these intermediate size molecules can behave quantum mechanically with respect to their translational motion. A value for the moment of inertia of a fullerene buckyball ($C_{60}$) has been referred to in the literature as $1.0 \times 10^{-43}$ kg·m$^2$; additional measurements have been reported for other fullerenes.[15] The moment of inertia values for these medium-sized molecules that have been shown to exhibit superposed quantum states are roughly 26 orders of magnitude smaller than the quantum-classical boundary estimated above in Eqn. (2). This extensive range of magnitudes for moments of inertia that are above those of the fullerenes while still well below the quantum-classical threshold value suggests that chemical structures considerably larger than fullerenes should also exhibit quantum interference effects, according to this criterion.

Macromolecules have somewhat larger moments of inertia than the fullerenes; that of the medium-sized protein lysozyme is reported to be $5 \times 10^{-41}$ kg·m$^2$; while a macromolecular assembly, a ribosome, has a reported moment of inertia five orders of magnitude larger; and a tobacco mosaic virus has a reported moment of inertia seven orders of magnitude larger.[16]

Most viruses that have been studied have diameters between about 10 and 300 nanometers.[17] Because of the wide range in their sizes, viruses might also be expected to exhibit a spread of values for their moments of inertia, with values perhaps more typically of the order of magnitude of $10^{-33}$ kg·m$^2$ to $10^{-35}$ kg·m$^2$; as noted above, the moment of inertia of a tobacco mosaic virus is in this range. Moments of inertia such as those estimated or measured for viruses are well within the range of values associated with quantum behavior for entire objects according to the present criterion, and so viruses would be expected to behave in a quantum manner unless brought into classicality by other effects.[6,9-11]

Romero-Isart and colleagues consider as an initial case the possibility of working with common influenza viruses, objects of size about 100 nanometers.[1] They also consider working with tobacco mosaic viruses, which have rod-like shapes with widths of about 50 nm and lengths of almost a micron.[1] As noted above, according to the present criterion, both the common influenza virus and the tobacco mosaic virus have moments of inertia that are far below the threshold of obligatory classical behavior. Hence, these viruses could be within the range of possible quantum behavior unless brought into classicality by other effects.[6,9,10]

Romero-Isart and colleagues also consider the possibility of conducting these experiments with tardigrades, small segmented eight-legged animals which range in size from about 100 μm to about 1.5 mm.[1] Taking the infamous "assume a spherical cow" approach, we can estimate moment of inertia values for typical tardigrades. In the case of tardigrades, the moments of inertia so estimated would be close to or above the estimated value for the threshold moment of inertia. Therefore, on the basis of this criterion, tardigrades would be classified as most probably classically behaved as entire creatures. Consequently, we might expect that it would be far more difficult if not impossible to observe these organisms in superposed quantum states.

**Summary and discussion**

To summarize, it appears that the experiments that have been proposed to create and observe quantum superposition of small living organisms may be expected to succeed in the cases they have discussed involving influenza viruses and tobacco mosaic viruses, but that the success of the corresponding experiment for the case of tardigrades remains doubtful because of the magnitude of the size and mass or moment of inertia of these small creatures.

We note that quantum objects can be brought into classicality by effects other than the limitations imposed by the expansion and finite lifetime of the universe that has been the basis for our analysis. Notably, decoherence effects and related effects dependent on interactions with the local environment can bring about classical behavior in objects that would otherwise behave quantum mechanically.[9-11]

It should be noted that the general criterion distinguishing classical from quantum behavior that is given in Eqns. (1) and (2) was derived for and hence expected to be applicable to free objects, whereas these experiments involve objects trapped inside cavities, so while the conclusions drawn from comparison with this criterion may be valid and certainly are of interest, its application under these circumstances might not be fully justified.

We also note that the estimates for a critical threshold value separating classical from quantum behavior are fairly crude estimates, good to an order of magnitude at best, and that more careful and systematic work examining the behavior of quantum wave functions in an expanding universe would be desirable in order to address potentially fundamental limitations on the feasibility of experiments such as these in a more thorough manner.

quantumlivingsizeeffects.doc
12 December 2009 draft